\documentclass[12pt,preprint]{aastex}

\def\deg {$^\circ$}

\slugcomment{Accepted for publication in Jan.~20, 2005 issue of ApJ}
\shortauthors{K.\ Gutierrez et al.}
\shorttitle{A~115 and 3C~28.0}
\begin{document}
\title{The Off-Axis Galaxy Cluster Merger Abell 115}
\author{K. Gutierrez,\altaffilmark{1,2} H. Krawczynski\altaffilmark{1}}
\altaffiltext{1}{Washington University in St. Louis, Campus Box 1105, 
  1 Brookings Drive, St. Louis, MO 63130}
\altaffiltext{2}{Contact: mrground@hbar.wustl.edu}
\begin{abstract}
We have used a 50.4 ksec Chandra observation to examine the galaxy cluster
Abell~115 ($z$=0.1971). The X-ray surface brightness distribution shows a 
merger of 2 sub-clusters with substantial off-axis motion.
The northern sub-cluster moves toward the southwest and exhibits a strong 
X-ray surface brightness gradient toward this direction as well as a ``fan'' 
of X-ray bright Intra-Cluster Medium (ICM) toward the northeast. 
The southern sub-cluster shows a pronounced northeast-southwest elongation, 
probably as a consequence of the ongoing merger. 
The temperature of the ICM is highly non-uniform across the cluster. 
Although the merger is well underway and has had a substantial influence on 
the morphology of the 2 sub-clusters, it has apparently not disturbed the 
cool sub-cluster cores: the cores show projected temperatures of $\sim$3~keV 
and are substantially cooler than the surrounding plasma of $\sim$5~keV. 
The ICM in between the sub-clusters has a significantly elevated temperature 
of $\sim$8~keV. However, owing to the low surface brightness, we do not 
find direct evidence for shocks in this region.
The X-ray surface brightness distribution of the northern sub-cluster is highly 
non-uniform. The data show a bright central core region of about 10'' (30~kpc) diameter.
A highly significant surface brightness discontinuity separates the core from the 
ambient ICM. We discuss several possibilities to explain the compact core, and its 
relation to the central radio galaxy 3C 28.0 and the sub-cluster's cooling flow.
Further outwards, at about 24.6'' (80 kpc) to the southwest of the sub-cluster core, 
the surface brightness distribution shows a jump associated with a contact discontinuity. 
The Chandra imaging spectroscopy data do not show evidence for a shock associated with 
the surface brightness jump. We conclude that it is associated with denser 
plasma moving sub-sonically through the ambient plasma. 
\end{abstract}

\section{Introduction}

\label{Gutierrez:intro}
%
%
Observations of cluster mergers allow us to witness the formation of large, 
gravitationally bound systems, and thus to study a key aspect of the process
of structure formation in our universe. The mergers result in the formation 
of large virialized systems, and play an important role 
in the ``eco-systems'' consisting of dark matter halos, Intra-Cluster 
Medium (ICM), galaxies, stars, and super massive black holes. 
Although the relative contributions of cluster mergers, galactic wind shocks, 
cluster accretion shocks, and outflows from Active Galactic Nuclei (AGN) 
are still somewhat uncertain, cluster mergers are believed to dominate the 
heating of the ICM. A detailed study 
of mergers is therefore essential to explain the galaxy and star formation 
efficiency in galaxy clusters, and to account for $\sim$80\% of the 
baryonic cluster mass residing between galaxies rather than inside galaxies.
A comprehensive summary of the astrophysics of cluster merger has recently
been given by Sarazin (2002).

%
%
There are three main observational channels to study cluster mergers: 
(1) Optical spectroscopy gives the location and line of sight 
velocities of member galaxies; (2) X-ray observations can be used to 
determine the temperature, density, and chemical composition 
of the hot ICM that fills the space between galaxies; 
(3) gravitational lensing observations yield constraints about the 
dark matter distribution, complementary to those inferred from 
X-ray observations. 
%
%
The Chandra and XMM-Newton X-ray telescopes can measure the spatial 
variation of the ICM temperature, density and chemical composition. 
With its unprecedented angular resolution of 0.5'', Chandra images 
have revealed clear evidence for shocks and Mach cones 
\cite{Mark:02,Kemp:04}, as well as for plasma discontinuities not 
associated with shocks, so-called cold-fronts \cite{Mark:00}. 
The ICM shock heating has thus become directly observationally accessible.
%
%

Besides cluster merger, another topic which we will discuss 
is the interaction of AGN with the ICM. In more relaxed clusters, 
rapid radiative cooling of the ICM near the cluster cores should result 
in ``cooling flows'' of the ICM toward the centers of the clusters \cite{Fabi:94}. 
At the cluster centers cooled ICM should either form cold clouds or 
undergo gravitational collapse and form stars. 
While the non-detection of such cold clouds or avid star formation 
had been a long-standing inconsistency of the cooling flow picture, 
the Chandra and XMM-Newton non-detections of cold ICM ($k_{\rm B} T$ $<$ 1.5 keV) 
at the centers of cooling flow clusters (e.g.~Peterson et al.\ 2003) 
has recently stirred considerable 
debate (see Donahue \& Voit 2003 and references therein).
Various authors proposed that bubbles of radio plasma ejected by  
AGN at the cluster centers might counteract the ICM cooling \cite{Boeh:02,Chur:02}.
The first bubbles were detected with ROSAT in the Perseus 
cluster \cite{Boeh:93} and in Cygnus A \cite{Cari:94}. Chandra observations 
have revealed radio bubbles in a large number of clusters (Blanton, 2003, 
Birzan, 2004 and references therein).
Detailed images of bubbles in the Perseus \cite{Fabi:03} and Virgo \cite{Form:04} 
clusters show very complex surface brightness distributions with 
evidence for contact discontinuities, shocks, and even sound waves. 
During the early phase of bubble formation, supersonic inflation of the 
bubble may produce a bow shock which heats the ICM (see e.g.~Ruzkowski et~al 2004).
If the ICM is sufficiently viscous, bubbles may be inflated subsonically, 
and heat the ICM by viscous dissipation \cite{Reyn:04,Ruzk:04}. 
Other proposed heating mechanisms invoke the p$\,\cdot\,$dV work done by the 
inflating bubble on the ICM, or mixing of inner cool and outer hot 
ICM induced by rising bubbles.

%
%
In this paper, we report on Chandra observations of the cluster
Abell 115 (A~115), a richness class~3 galaxy cluster at 
$z$=~0.1971 \cite{Abel:89}.
{\it Einstein} observations with the Imaging Proportional Counter 
(IPC, 2.6 ksec) and the High Resolution Imager (HRI, 8.2 ksec) 
revealed a double peaked X-ray surface brightness \cite{Form:81}. 
The cluster has a large X-ray luminosity of 
4.7$\times 10^{45}\,$$\rm erg\, s^{-1}$. A surface brightness 
deprojection analysis indicated cooling flows of 
313$^{+360}_{-186}$ $M_\odot$ and $>$112 $M_\odot$ in the northern and 
southern sub-clusters, respectively \cite{Whit:97}.
ASCA X-ray observations of A~115 (32.0 ksec + 33.1 ksec)
indicated significant variation of the 
F(2~keV-10~keV)/F(0.5~keV-2~keV) hardness ratio, 
consistent with an ICM 
temperature of the northern component, 
the interface region, and the southern component of 4.9~keV, 11~keV, 
and 5.2~keV, respectively \cite{Shib:99}. 
The cluster has extensively been studied in the 
optical \cite{Noon:81,Beer:83,Rako:00,Pink:00}. 
The northern and southern X-ray components are associated with 
dense regions of cluster galaxies; a third galaxy concentration 
is found in a region of rather low X-ray surface 
brightness in the east of the cluster \cite{Beer:83}.
The northern sub-cluster harbors the very strong radio galaxy
3C~28.0 \cite{Rile:75,Mack:83,Fere:84,Giov:87,Owen:97,Govo:01}.
Its host galaxy is the Brightest Cluster Galaxy (BCG) of the 
northern cluster component. The radio morphology is intermediate 
between the classical Fanaroff-Riley I and II classes.
Its 178 MHz luminosity of 
$L_{\rm 178\,MHz}=\,$2.3$\times\,$$10^{26}\,$$\rm W\,Hz^{-1}\,sr^{-1}$ 
is 10 times higher than usual for FR~I sources.

%
%
We proposed A~115 observations with the Chandra X-ray telescope
to scrutinize the cluster merger for temperature inhomogeneities and
to search for merger shocks. As an additional motivation, an archival 
ROSAT HRI (29~ksec) image showed indications for a morphological correlation
between the radio brightness of 3C~28.0 and the ICM surface brightness.
In this paper, we describe the results of the 50.4 ksec Chandra observation.
We describe the data 
cleaning and analysis in Sect.~\ref{Gutierrez:Observations}. 
We present the surface brightness and temperature distributions
in Sects.\ \ref{Gutierrez:morphology} and \ref{Gutierrez:temp}, respectively.
We discuss surface brightness discontinuities of the northern sub-cluster (A~115~N)  
and their interpretation in Sect.\ \ref{Gutierrez:northern}.
We scrutinize the morphological correlation of the radio galaxy 3C~28.0 and
the X-ray surface distribution in Sect.\ \ref{Gutierrez:Radio}.  
In Sect.\ \ref{Gutierrez:discussion} we summarize the results.

In the following we use H$_{0}=h_0 \times 100$ km s$^{-1}$ Mpc$^{-1}$, 
$h_0=0.71$, $\Omega_{\rm M}\,=$ 0.27, and $\Omega_{\rm vac}\,=$ 0.73, 
putting the cluster at a luminosity distance of
955 Mpc and a angular distance of 666 Mpc;  1'' corresponds to 3.23 kpc.
Statistical errors are given on 90\% confidence level.  

\section{Data Preparation and Analysis}
\label{Gutierrez:Observations}
The Chandra ACIS~I observation was taken on October 7-8, 2002 
(Chandra ObsId 3233). The observation used the chips ACIS I0-I3 as 
well as ACIS S2. The data were taken in ``Very Faint'' (VF) mode to 
minimize background counts and the ACIS S3 chip was turned off to 
keep the telemetry rate low during background flares. 

The data were analyzed with CIAO~3.0.1 analysis 
software\footnote{Information about CIAO and all the analysis steps described
in this paper can be found at: http://cxc.harvard.edu/ciao/} 
using the CALDB~2.26 calibration 
data base\footnote{http://cxc.harvard.edu/caldb/}.
After removing bad pixel events, the data were screened for background flares, 
but none were found. We corrected the data for the Charge Transfer 
Inefficiency (CTI) with the 
script \textit{acis\_process\_events} and the option \textit{apply\_cti=yes}. 
We used \textit{apply\_gain} 
to correct for the secular change in the ACIS I 
gain\footnote{http://cxc.harvard.edu/ciao/threads/acistimegain/}.
The VF-specific filter algorithm 
reduced the background by 30\% while retaining 98\% of the signal.
We used the standard method to create a background data set from 
ACIS~I Blank Sky data using the script \textit{make\_acisbg}.
We applied the same VF filter to the background data set as to the 
cluster data.
We produced flat-fielded images by correcting the 0.5-1 keV, 1-2 keV, 
and 2-5 keV images with the appropriate exposure maps, and subsequently 
combining the 3 images with a spectral weight.
We constructed a hardness ratio map by using the flatfielded and background
subtracted (2-5 keV) and (1-2 keV) surface brightness maps.
After smoothing both maps in the same way, we divided them to give the 
hardness ratio map.

The spectral analysis was performed with \textit{sherpa} and used 
the script \textit{acisspec}.
The script can handle extended sources and accounts for the different 
exposure times of the data file and the background file.
The spectral analysis uses only events from 0.5 keV to 6 keV.
Above 6 keV the signal to noise ratio is poor; below 0.5 keV the 
response of the instrument is plagued by systematic uncertainties.    
For the spectral analysis, we scaled the counts below 1.8 keV by the
empirical factor of 0.93 to correct for the quantum efficiency degradation in
front illuminated chips (M.\ Markevitch, private communication). 
The energy spectra were binned with the requirement to have at 
least 5 counts per bin. The regions we used to extract energy spectra 
avoided detected point sources.

The study of the morphological correlation between the ICM surface brightness 
distribution and the radio emission of 3C~28.0 used the 1.4 GHz radio map 
by Feretti et~al.~(1984). 
%
\section{Results}
\label{Gutierrez:results}
\subsection{Overall Morphology of the Binary Merger}
\label{Gutierrez:morphology}
\begin{figure}
\plotone{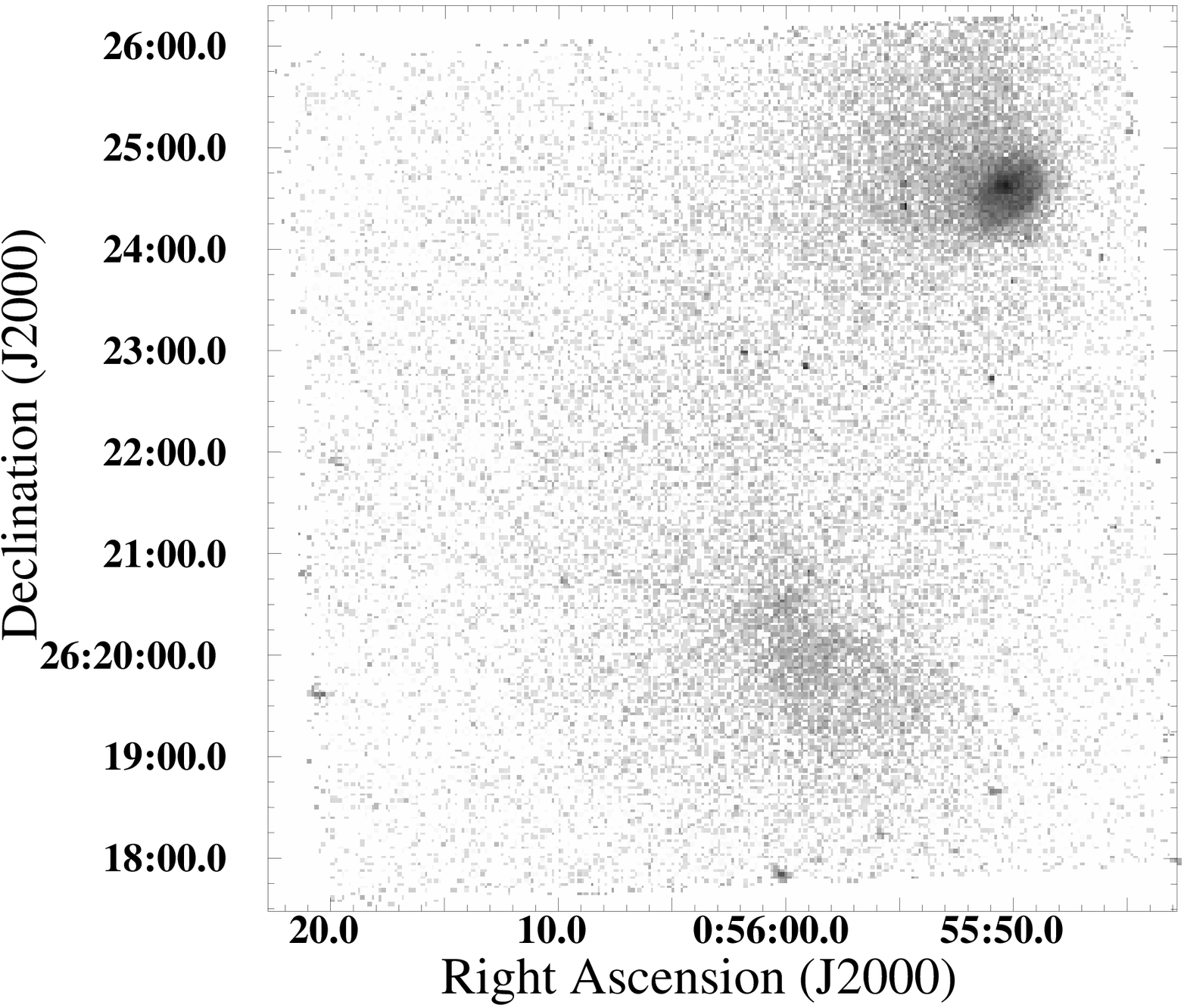}
\caption{The surface brightness distribution (0.5 keV - 5 keV) of the galaxy
cluster A~115. Sub-structure in the northern sub-cluster can clearly be recognized.
The data has been flatfielded and ``fluxed''. Lightest gray
corresponds to $4 \times 10^{-8}$ photons cm$^{-2}$ s$^{-1}$ arcsec$^{-2}$, and black to 
$4 \times 10^{-6}$ photons cm$^{-2}$ s$^{-1}$ arcsec$^{-2}$. This image spans 8.9' $\times$
8.9' (1.72 Mpc $\times$ 1.72 Mpc).}  
\label{Gutierrez:cluster}
\end{figure}

Fig.~\ref{Gutierrez:cluster} shows the X-ray surface brightness distribution of A~115.
The overall morphology shows a merger with very substantial off-axis motion
and roughly resembles the well known morphology seen in optical images of 
off-axis galaxy mergers, e.g.~optical images of the Antenna galaxy merger.
The northern sub-cluster exhibits a wedge like shape and 
appears to be moving toward the southwest direction. 
The sub-cluster's surface brightness distribution shows a strong gradient 
at its southwestern side and an X-ray bright ``fan'' toward the northeast. 
The core of the radio galaxy 3C~28.0 is roughly centered on the brightest X-ray 
region of A~115~N.

The merger skewed the southern sub-cluster (A~115~S), resulting in a pronounced 
elongation in the northeast to southwest direction. 
We detected point-like X-ray emission from a spectroscopically verified cluster galaxy 
that lies at the northeastern edge of the bright core of the southern sub-cluster. 
The location suggests that this galaxy is the central galaxy of the southern 
sub-cluster. 
In between the 2 sub-clusters, offset toward the east direction, the Chandra image 
shows faint X-ray emission from a region that connects the 2 sub-clusters.
The absence of strong surface brightness gradients do not allow us to determine 
unambiguously if A~115~S moves into the northeast or southwest direction. 
A possibility is that the sub-cluster moves toward the northeast, with the
dark matter core and the central galaxy leading and the ICM trailing behind.

\begin{figure}
\plotone{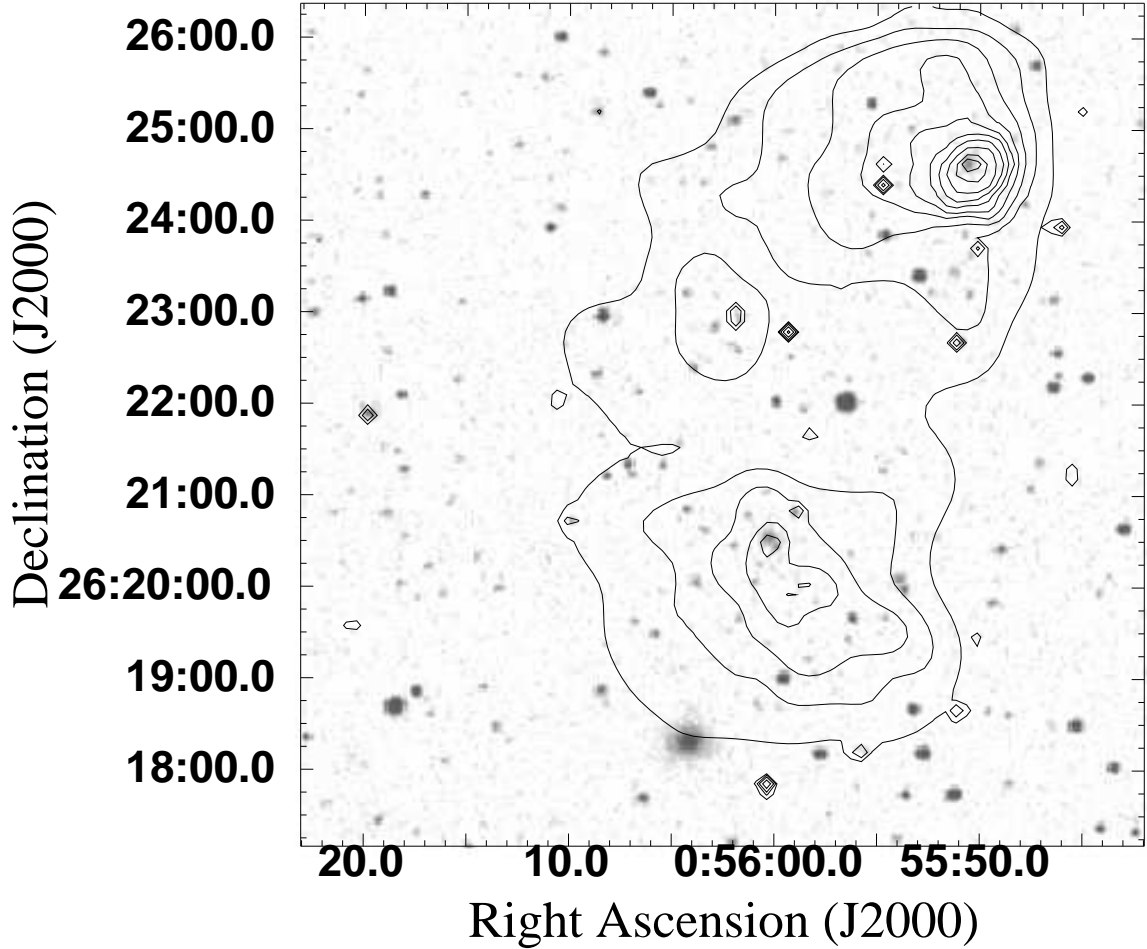}
\caption{Optical image from the STScI DSS survey (grayscale)
and Chandra surface brightness contours to aid the eye.
For both sub-clusters, the peak of the X-ray surface brightness coincides with
an optically detected galaxy.  This image spans 9.3' $\times$
9.3' (1.81 Mpc $\times$ 1.81 Mpc).}
\label{Gutierrez:optical}
\end{figure}

\begin{figure}
\plotone{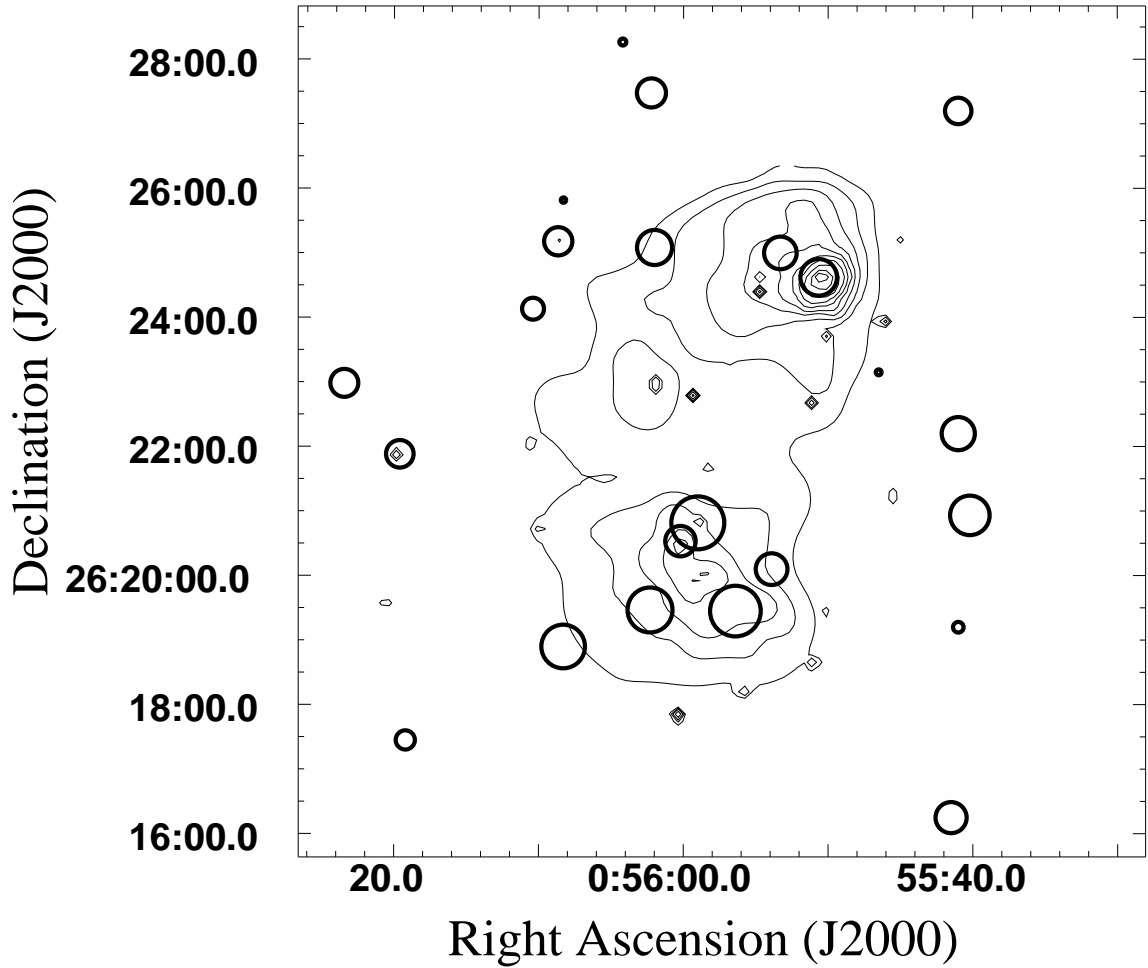}
\caption{Spectroscopically verified cluster galaxies represented as circles with 
Chandra surface brightness contours to aid the eye. The circle diameters encode the redshifts of the 
galaxies, higher redshifts corresponding to larger circles.
The lowest (highest) redshift shown is 0.180 (0.202). This image spans 13.3' $\times$
13.2' (2.58 Mpc $\times$ 2.56 Mpc).}  
\label{Gutierrez:redshift}
\end{figure}

Fig.~\ref{Gutierrez:optical} shows the Chandra contours overlaid on the optical Digital 
Sky Survey\footnote{http://archive.stsci.edu/cgi-bin/dss\_form} (DSS) image.
Several galaxies can be recognized that emit in the optical and in the X-ray band.
For both sub-clusters, the peak of the X-ray surface brightness coincides with a 
galaxy detected in the optical. Fig.~\ref{Gutierrez:redshift} shows the location and
redshifts of all spectroscopically verified cluster galaxies that are listed in the 
Simbad astronomical data base\footnote{http://simbad.u-strasbg.fr/Simbad}.
The radii of the circles are encoded with the redshifts of the sources -- small circles
representing low redshifts, and large circles representing high redshifts.
The redshifts are from Beers et~al.~(1983), Pinkney et~al.~(2000), and Zabludoff et~al.~(1990).  
The majority of  galaxies in A~115~S seem to recede faster from us than the average A~115~N galaxies.
A notable exception is the galaxy at or near the peak of the A~115~S X-ray surface distribution.
The average velocity difference between the 3 galaxies close to the northern sub-cluster core
and the 6 galaxies close to the southern sub-cluster core is $0.003c\,=$ 900 km s$^{-1}$, which 
is below the clusters escape
velocity of $0.005c$.  The different average redshifts of the galaxies in the 2 sub-clusters
indicate that the orbital motion is not fully contained in the plane of the sky. 

Fig.\ \ref{Gutierrez:RadioHalo} shows the Chandra contours overlaid on the 1.4~GHz 
radio brightness map of Govoni et al.\ (2001).
The deep radio map shows 3C~28.0, as well as an additional source toward the northeast of the cluster.
A 10' (2 Mpc) long radio arc connects the two radio sources and extends further towards the northeast
of the cluster. Govoni et al.\ discussed the origin of the radio emission and emphasized that the 
orientation and location of the arc argued against a cluster relic.
In view of the rapid movement of A~115~N towards the southwest, as evident from the
Chandra data, the radio emission may be interpreted as tails of radio plasma 
trailing the radio galaxies. 

A search for X-ray point sources near A~115 resulted in five interesting detections.
Four sources are associated with cluster galaxies, 
and one source with the infrared source IRAS F00536+2611.
Further to the west (9' from A~115~N) we find a very strong X-ray
source with 948 Chandra counts and without any counterparts in the Simbad data base.
This source lies at $\alpha$(J2000)$=$ 0:55:09.2 $\delta$(J2000)$=$ +26:27:14 
and is designated here CXOU J005509.2+262714.  The source exhibits a very hard energy 
spectrum and is not well fitted by a one-component Raymond-Smith model. 
A simple power-law fit gives a photon index of $\Gamma=0.93^{+0.05}_{-0.05}$, 
and a 1 keV flux amplitude of $P_{amp}=3.01^{+0.11}_{-0.11} \times 10^{-5}$ 
photons keV$^{-1}$ cm$^{-2}$ s$^{-1}$ with $\chi^{2}(dof)\,=$ 42.6(43). 
We do not find evidence for flux variability during the 50 ksec Chandra
observation.   Optical STScI DSS show an optical counterpart, however 
no radio counterpart is seen in the 1.4 GHz radio map of Govoni et al.  
The source is close to a chip boundary and
appears to be extended, however using the ChaRT\footnote{http://cxc.harvard.edu/chart/}
simulator we obtained an almost shaped image from a point source.  
We therefore conclude that this source is most likely a point source.
All the point sources are listed in Table~\ref{Gutierrez:t2}.

\subsection{Temperature and metallicity of the ICM}
\label{Gutierrez:temp}
\begin{figure}
\plotone{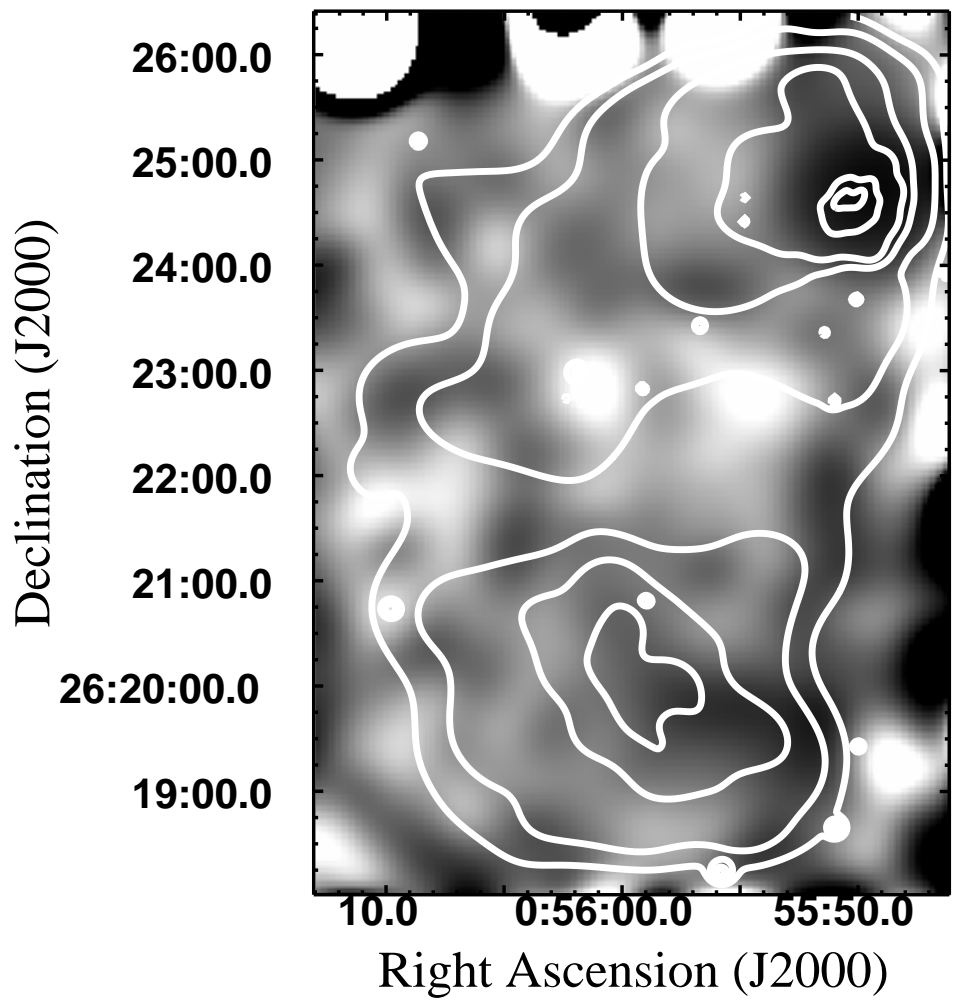}
\caption{(2-5 keV)/(1-2 keV) hardness ratio map (grayscale) with Chandra surface brightness contours to aid the eye. 
The light (dark) regions show regions with very hard (soft) energy spectra. 
The cool cluster cores and the hot 
gas between the sub-clusters can be recognized. 
The hardness ratios vary from 0.9 near the sub-cluster cores to 2.1 in-between the galaxies.
The circular regions on the top and right hand sides are artifacts 
from statistical fluctuations near the chip boundaries. Smoothing with Gaussian of 37'' FWHM.
This image spans 8.4' $\times$ 6.1' (1.63 Mpc $\times$ 1.18 Mpc).}
\label{Gutierrez:HardnessRatio}
\end{figure}

\begin{figure}
\plotone{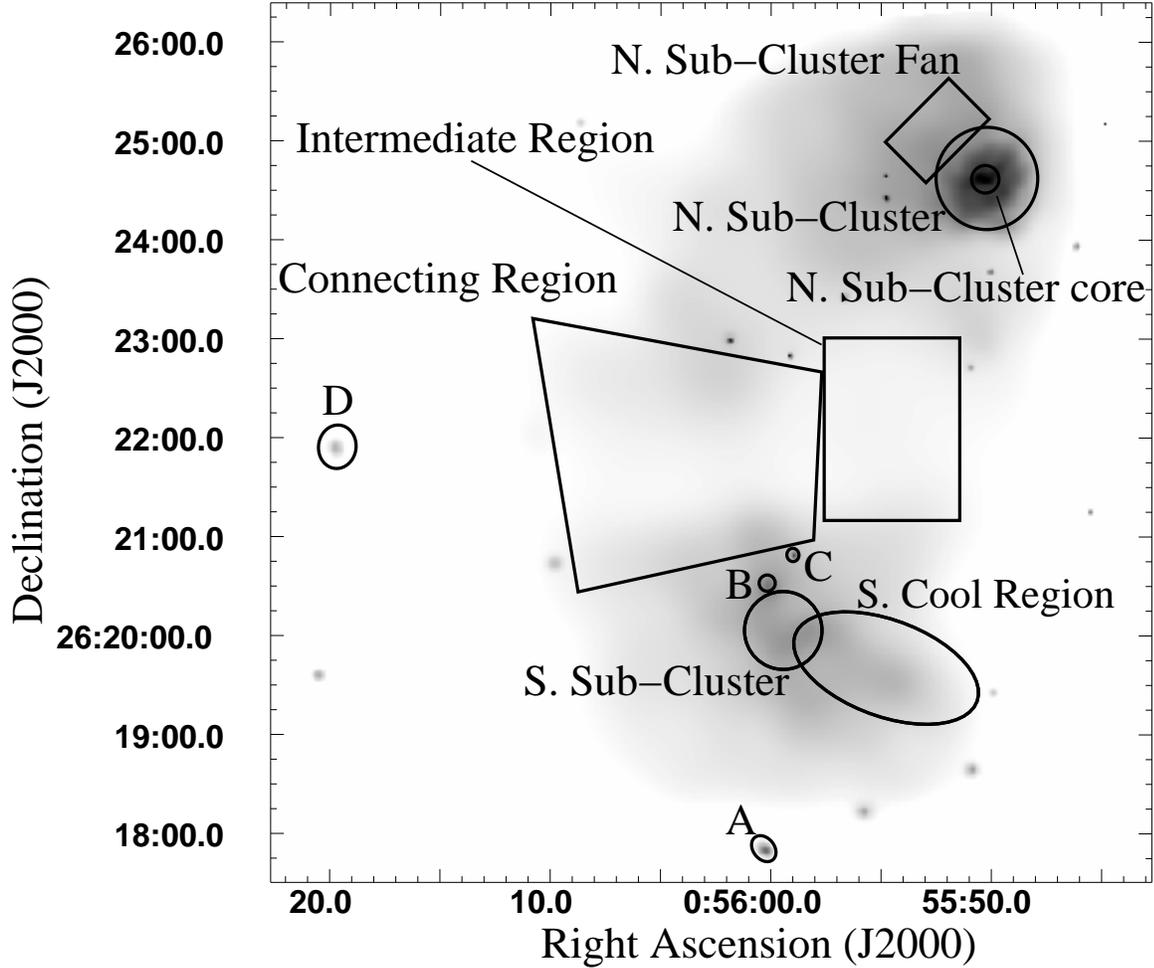}
\caption{The image shows the regions used for the spectral analysis 
on top of a smoothed surface brightness map of the cluster. 
The results of spectral fits are given in Table~\ref{Gutierrez:t2}. 
We marked 4 point sources with circles/ellipses and the letters A to D
(see Table~\ref{Gutierrez:t1}). 
Sources E and F from Table~\ref{Gutierrez:t1} lie outside the region of the 
sky shown in this image. The grey-scale image was smoothed with the 
\textit{csmooth} algorithm. This image spans 9' $\times$ 9' (1.74 Mpc $\times$ 1.74 Mpc). } 
\label{Gutierrez:region}
\end{figure}

The (2-5 keV)/(1-2 keV) hardness ratio map of the X-ray emission is shown in Fig.~\ref{Gutierrez:HardnessRatio}.
The two maxima in the surface brightness distribution show up as regions of rather cold ICM.
Between the two sub-clusters is a region of very hot ICM. We ran a full spectral analysis for the 
most interesting regions, shown in Fig.~\ref{Gutierrez:region}. 
The results from one component Raymond-Smith model fits are listed in Table~\ref{Gutierrez:t1}.
The temperature of the two sub-clusters is lowest near their cores. For the northern and the southern
sub-clusters we get projected temperatures of 2.2~keV and 3.7~keV, respectively. 
The metallicity is highest ($\sim$0.4 solar) near the core of A~115~N. The rest of the cluster
exhibits a metallicity close to 1/5th solar.
For the thin region connecting the two sub-clusters and for the region directly between the 2 sub-clusters
we get temperatures of 7.2 keV and 8.7 keV, respectively. Thus, the Chandra observation confirms the 
earlier result by Shibata et~al.~(1999) of hot regions between the 2 sub-clusters based on ASCA observations. 
Based on the luminosity-temperature correlation, the emission weighted mean temperatures of 
A~115~N and A~115~S are 5.2 keV and 4.8 keV, respectively \cite{Whit:97}.
Thus, the sub-cluster cores are indeed relatively cold and the regions between the sub-clusters 
are relatively hot.
The Chandra image does not show direct evidence for a shock or a shock system between the two 
sub-clusters that might have caused the elevated ICM temperature.

\subsection{The structure of the northern sub-cluster}
\label{Gutierrez:northern}
\begin{figure}
\plotone{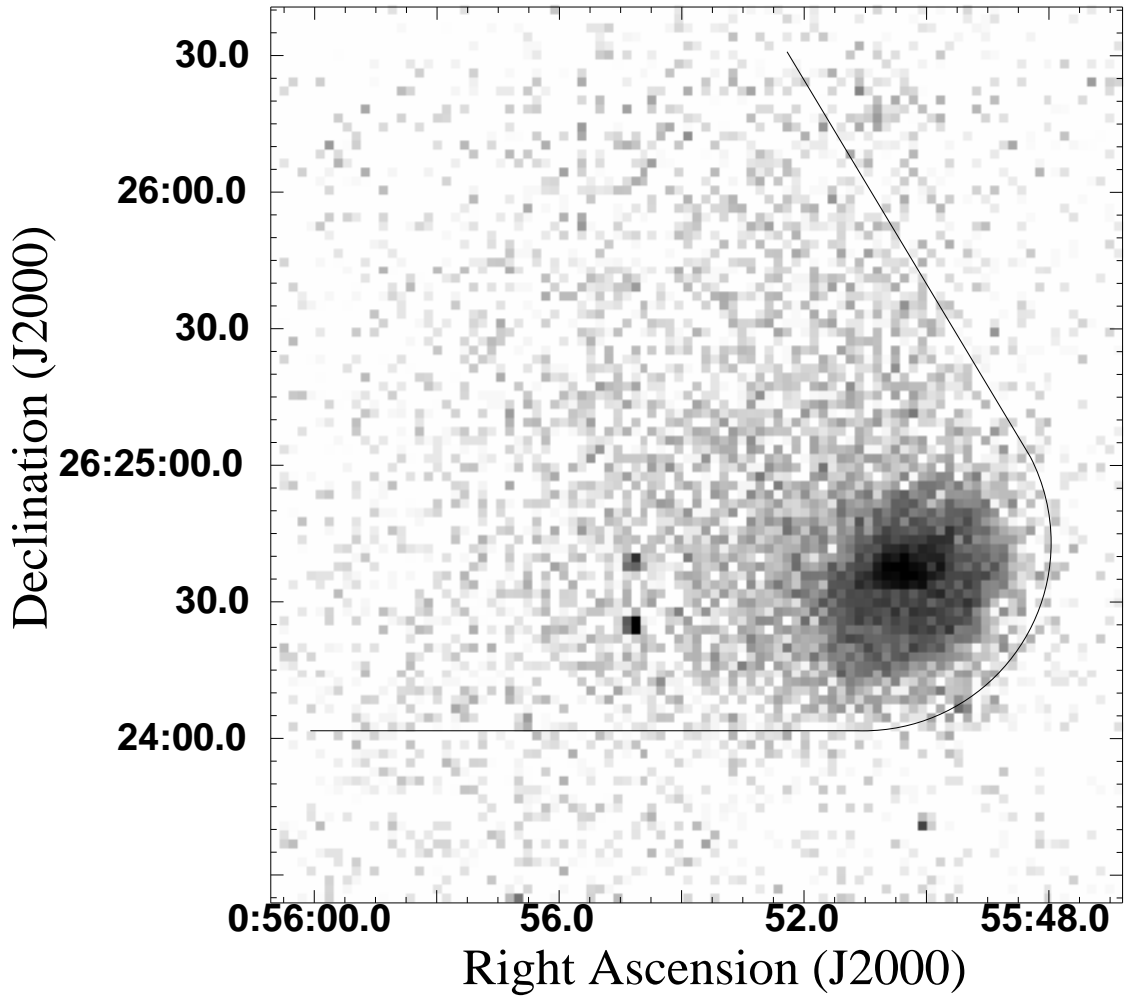}
\caption{Image of the surface brightness distribution (0.5 keV - 5 keV) of the northern 
sub-cluster. The solid line exemplifies the shape of the surface brightness
distribution. Although the shape resembles that of a Mach cone, the imaging
spectroscopy data do not show evidence for the presence of a shock
(Same grayscale as in Fig.\ \ref{Gutierrez:cluster}).  This image spans 3.3' $\times$ 3.1' (640 kpc $\times$ 600 kpc). }
\label{Gutierrez:north}
\end{figure}

\begin{figure}
\plotone{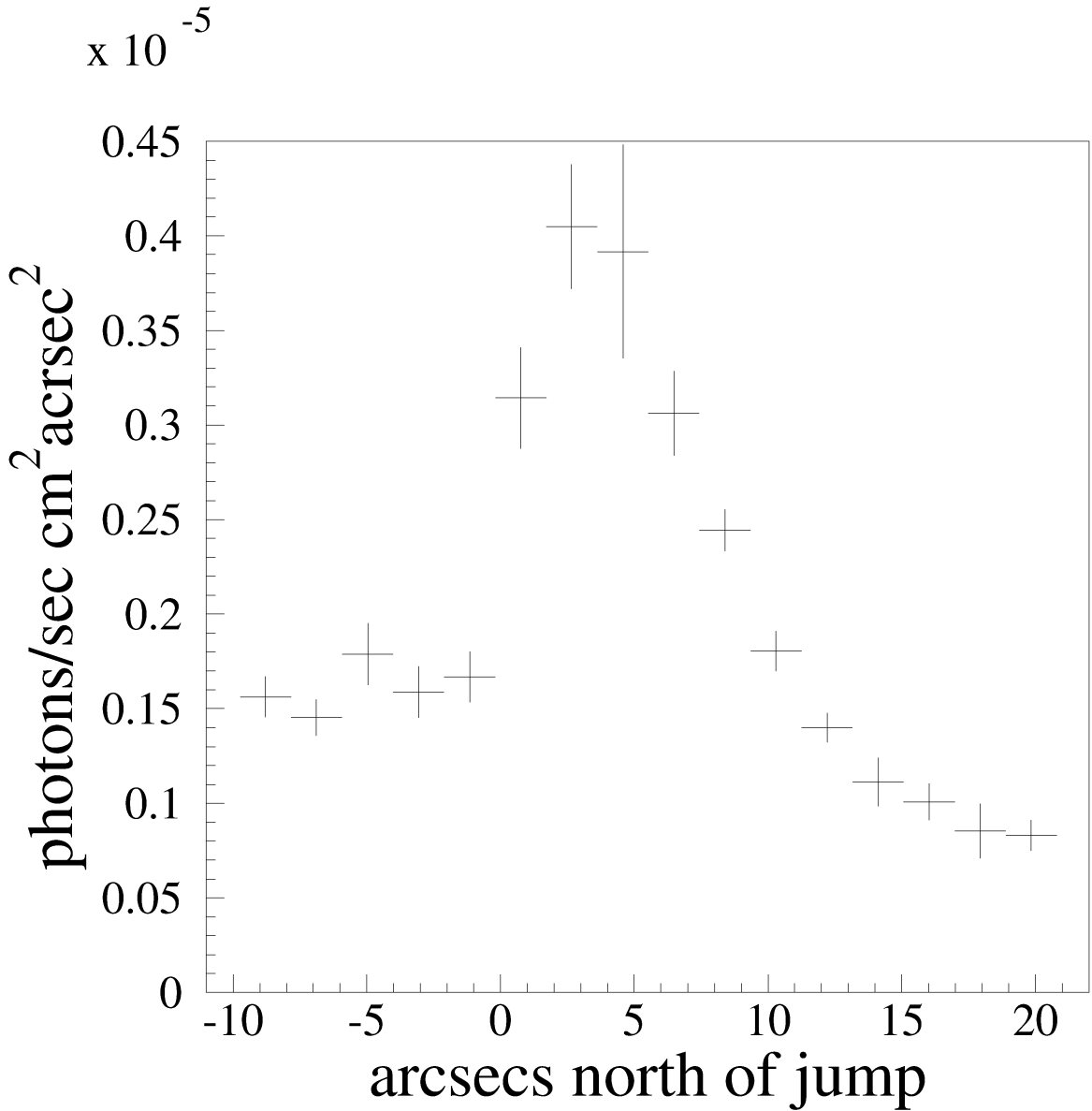}
\caption{Surface brightness profile of the northern sub-cluster from south to north.
Each bin contains the surface brightness averaged over 24 pixels ($\sim$ 12'') in east-west direction.
The X-ray bright core can clearly be recognized. A sharp surface brightness discontinuity 
is evident in the south of the sub-cluster. Towards the north, the shape of the bright 
region is more irregular, resulting in a more gradually changing surface brightness profile.}
\label{Gutierrez:line}
\end{figure}

\begin{figure}
\plotone{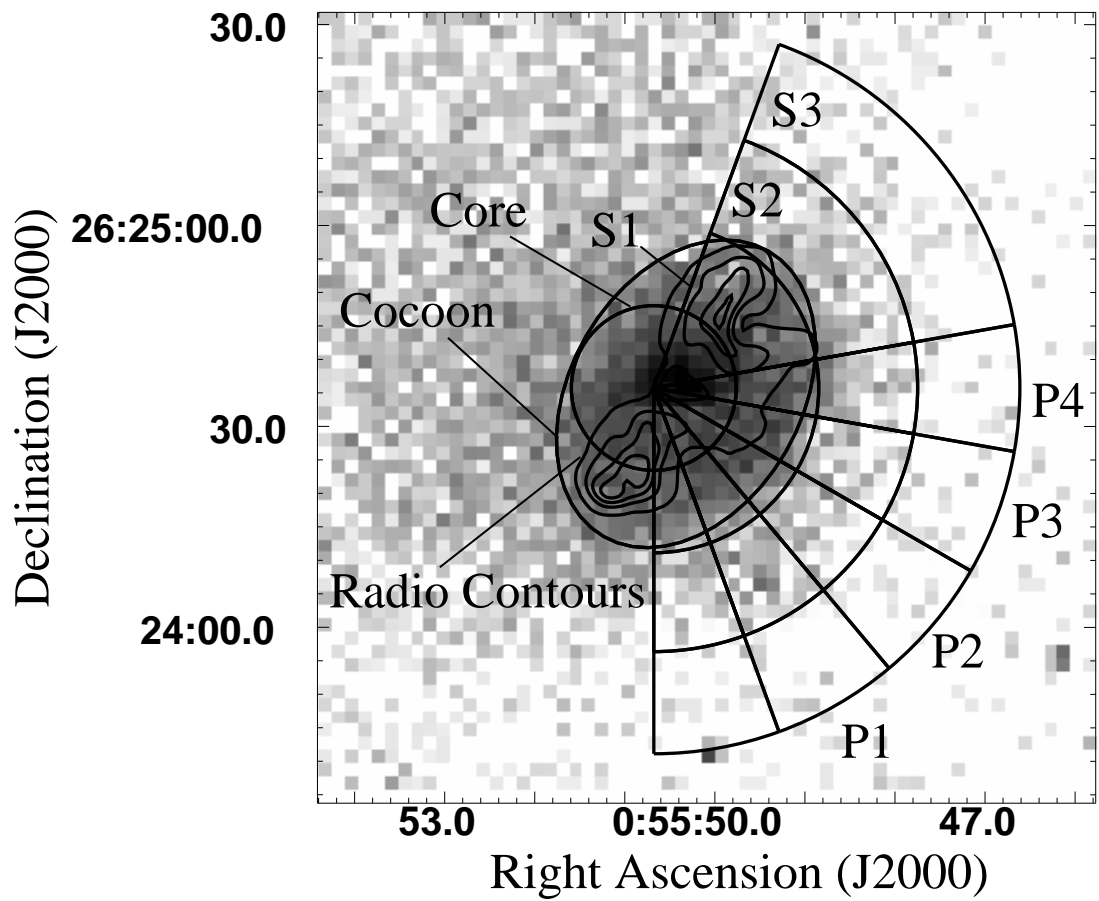}
\caption{Image of the central region of the northern sub-cluster.  
The Gray scale map shows the X-ray brightness and the contours 
show the 1.4 GHz radio brightness. 
The pie slices P1, P2, P3, and P4 have been used to extract the surface brightness
profiles in Fig.\ \ref{Gutierrez:f3}. The regions C, S1, S2, and S3 have been 
used for the deprojection analysis in Fig.~\ref{Gutierrez:f4}.  This image spans 2' $\times$ 2' (390 kpc $\times$ 390 kpc). }
\label{Gutierrez:core}
\end{figure}

In this section we scrutinize A~115~N. We start with the 
sub-cluster core, and subsequently work our way further outwards. 
In Fig.\ \ref{Gutierrez:north} one can recognize an X-ray bright region right at the core 
of the sub-cluster. As shown in the north-south surface profile in Fig. \ref{Gutierrez:line}, 
the boundary of the region is associated with a highly significant surface brightness jump.
The region is approximately 10'' (30~kpc) wide and 10'' (30 kpc) high. The projected size
suggests that the emission might be associated with the Inter-Stellar Medium (ISM) of the 
central cD galaxy. The photon statistics are not sufficient to measure a difference in
temperature and chemical composition between the bright compact region and 
the ambient ICM. We will discuss the hypothesis further in Sect.\ \ref{Gutierrez:discussion}.
The surface brightness distribution of the A~115~N shows a steep gradient to the 
southwest of the sub-cluster core. We studied this region with radial surface brightness
profiles and a spectroscopic deprojection analysis. 
The sky regions used for the following analysis are shown in Fig.\ \ref{Gutierrez:core}.

\begin{figure}
\epsscale{0.65}
\plotone{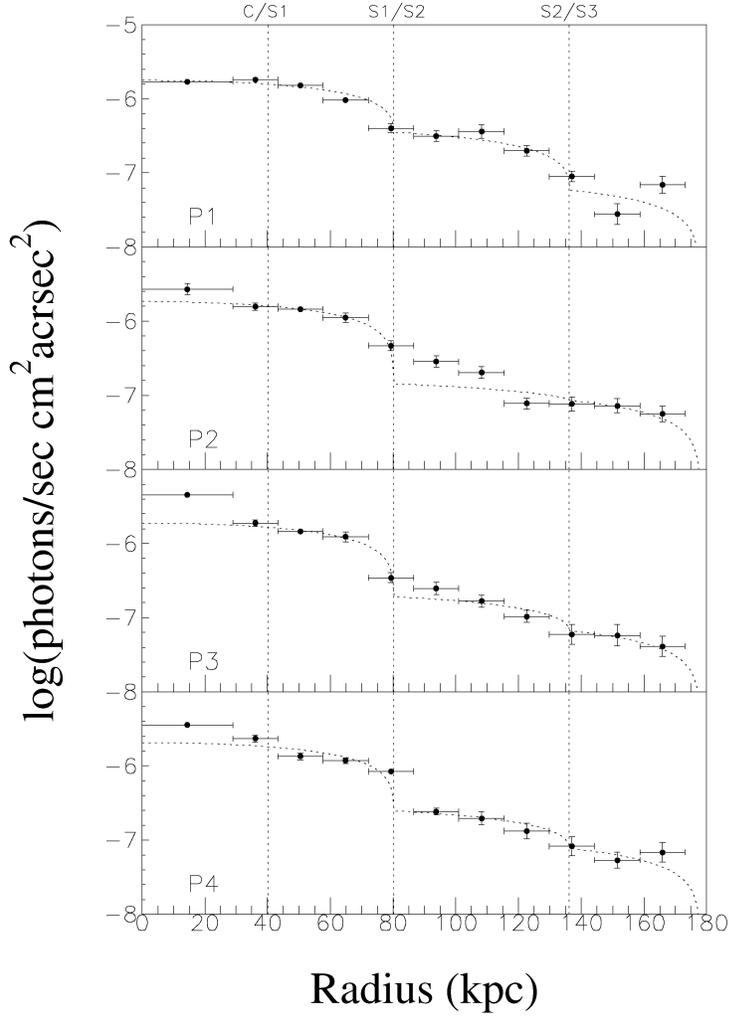}
\caption{ICM surface brightness profiles of 4 pie slices in the southwestern
part of the northern sub-cluster. The 4 slices P1-P4 are shown in 
Fig.~\ref{Gutierrez:core}. 
The dotted lines indicate the 4 regions described in the text, and used
for the deprojection analysis 
(C, S1, S2, S3, see also Fig.~\ref{Gutierrez:core}):
(C) a central sphere from 0 to 40 kpc; 
(S1) the first ICM shell from 40 kpc to 80 kpc; 
(S2) the second shell from 80 to 136 kpc; 
(S3) the third shell from 136 to 178 kpc. 
The curve shows a simple model prediction for the surface brightness that
assumes a sphere and 3 shells, each with ICM of constant X-ray emissivity. 
A jump in the surface brightness can be seen roughly at the C/S1 boundary.
The location of the jump depends slightly on the pie-section. 
For sections P1-P3 it can be found at 70 kpc core distance, 
and for section P4 at 85 kpc core distance.}
\label{Gutierrez:f3}
\end{figure}

\begin{figure}
\plotone{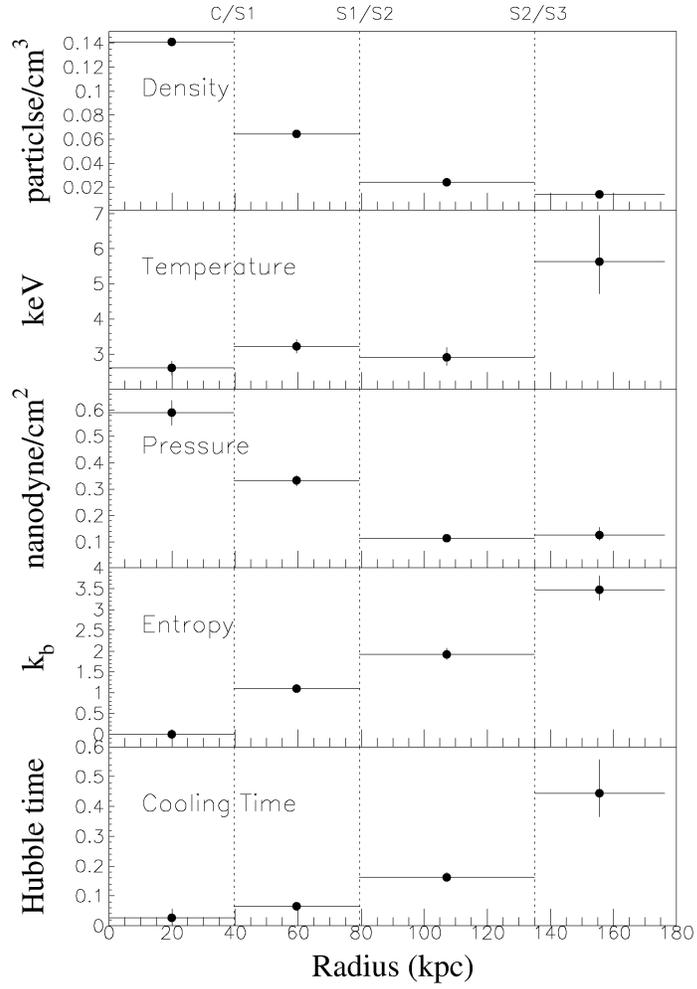}
\caption{Deprojected ICM particle density, temperature, pressure, entropy and cooling time 
profiles of the southwestern part of A~115~N
(from 20\deg east of north to due south, see the large fan shown in Fig.\ \ref{Gutierrez:core}).}
\label{Gutierrez:f4}
\end{figure}

Fig.~\ref{Gutierrez:f3} shows surface brightness profiles for 4 pie sections to the 
southwest of the core of A~115~N. The geometry of the 4 pie sections
(P1 - P4) are shown in Fig.\ \ref{Gutierrez:core}.
Close inspection of the profiles shows a jump at about 24.6'' (80 kpc) from the cluster core,
or more correctly, from the core of the radio galaxy 3C~28.0, which we chose as center for this study.
The location of the jump depends slightly on the pie-section. For sections P1-P3 it can be found
at 75 kpc from the core, and for section P4 at 85 kpc. For pie sections P1 and P2 the surface brightness
jumps by a factor of 2.5. For sections P3 and P4 it jumps by a factor of 4.
The jump associated with the compact bright region (ISM) of the central galaxy can not be seen 
here clearly, owing to the irregular shape of that region.
In rough approximation, the surface brightness profiles can be described by a simple model using 
a sphere surrounded by three spherical shells, and assuming constant volume emissivity in the 
four volumes (see Fig.~\ref{Gutierrez:f3}, dashed line).
To further elucidate the origin of the surface brightness jumps, we assumed spherical symmetry 
in the southwestern part of the cluster and performed a spectroscopic deprojection analysis \cite{Kraw:02}. 
The model uses a spherical core (C) surrounded by 3 ICM shells (S1, S2, S3), as shown in Fig.\ \ref{Gutierrez:core}.
Working from the outermost ICM shell inwards, we determine the 1 component Raymont Smith plasma parameters. 
Hereby we take into account, if applicable, the contribution of outer shells. 
Fig.~\ref{Gutierrez:f4} shows from top to bottom the derived ICM particle density, temperature, 
pressure, entropy, and cooling time profiles.
The ICM density falls from 0.14 particles cm$^{-3}$ at the cluster core to 0.014 
particles cm$^{-3}$ at the third shell. We obtain a temperature of $\sim$3~keV at the core 
and the innermost 2 shells. Shell 3 is substantially hotter at $5.6^{+1.3}_{-0.9}$ keV. 
The pressure drops from $5.90^{+0.47}_{-0.47}\times 10^{-10}$ dyne cm$^{-2}$
at the cluster core to $1.26^{+0.29}_{-0.20}\times 10^{-10}$  dyne cm$^{-2}$
at shell 3. The entropy increases monotonically from the cluster core to shell 3, 
with a total change of $3.47^{+0.37}_{-0.27} k_{\rm b}$. 
The radiative cooling time is shortest at the cluster core being $0.026\times1/H_{0}$
$\approx$ 360 million years.
The deprojected plasma parameters argue strongly against a shock being responsible for the surface 
brightness jump at the S1/S2 boundary. The temperature stays rather constant across the boundary and
the entropy increases from shell S1 to S2, contrary to the expectations for a shock.
In a more quantitative analysis, we use the Rankine-Hugoniot relations that relate the upstream 
to the downstream plasma parameters:
\begin{equation}
\frac{P_1}{P_2}\,=\frac{(2\,M^2\,-\,1)\gamma+1}{\gamma+1}
\label{RH1}
\end{equation}
\begin{equation}
\frac{n_2}{n_1}\,=\frac{2\,/\,M^2\,+\,\gamma-1}{\gamma+1}
\label{RH2}
\end{equation}
where $P_1$, $n_1$, and $P_2$, $n_2$ are the pressure and particle densities in 
shells 1 and 2, respectively; $M$ is the shock's Mach number, and $\gamma\,=$ 5/3 
is the adiabatic index. Throughout the entire sub-cluster the radiative cooling times are 
long compared to the dynamical time scales of the relative motion of the sub-clusters. 
The assumption of an adiabatic shock is thus well justified.
Using that the ICM emissivity $\epsilon$ is roughly proportional to $\sqrt{T}$ $n^2$ together with
the ideal gas law, we derive that a surface brightness jump by a factor of $\sim$3 
would be accompanied by a temperature increase by a factor of 1.4
and an entropy increase by 0.04~$k_{\rm b}$ across the shock.
Contrary to these expectations, the deprojected temperature increases only by a factor of 1.11
and the entropy drops rather than increases from shell 2 to shell 1.
We conclude that the surface brightness jump is not caused by a shock but by 
dense ICM coasting through thinner ICM.

The surface brightness distribution of A~115~N exhibits an X-ray bright fan to the 
northeast of the cluster core.
To test if this feature corresponds to a Mach cone, we have chosen 
the shells of the deprojection analysis such that the S2/S3 boundary approximately 
connects to the fan boundaries (see Fig.\ \ref{Gutierrez:north}). 
If the feature was associated with the bow shock of the cluster core, we would expect 
that the temperature and entropy increase from shell S3 to shell S2. 
The data, however, show that both decrease. 
More quantitatively, we use the angle of $\theta\,=$ 70$^\circ$ between the boundaries of the fan 
to compute a corresponding Mach number:  

\begin{equation}
M=\frac{1}{sin(\frac{\theta}{2})}=1.74
\label{MC1}
\end{equation}

Using the Rankine-Hugoniot 
conditions from Eqs.\ \ref{RH1}-\ref{RH2}, we infer minimum density, pressure, and temperature 
jumps from shell 3 to shell 2 by factors of 2, 3.5, and 1.7, respectively. 
Similarly the  entropy would be also be expected to increase by 0.1 $k_{\rm b}$.
Such jumps are clearly not observed.
We conclude that the X-ray bright fan to the northeast of the 
cluster core is not associated with a shock, but is produced by ICM trailing behind the sub-cluster core.
\subsection{Correlation of X-ray surface brightness and the radio morphology of 3C~28.0}
\label{Gutierrez:Radio}
Note that the morphology of 3C~28.0 is very similar at the two
wavelengths where we have high resolution radio maps, 1.4 GHz \cite{Fere:84}, and 5 GHz \cite{Rile:75}.
The radio data show double-sided jets that have drilled 2 channels through the compact 
X-ray bright region at the sub-cluster core. The jets transition into radio lobes right when they 
leave the compact region. The radio lobes clearly avoid the compact region.
The Chandra data show no evidence for a cavity or bubble associated with the radio lobes.
%
\section{Summary and discussion}
\label{Gutierrez:discussion}
\subsection{Off-axis cluster merger}
\label{Gutierrez:OffAxis}
The Chandra observation shows a binary cluster merger with a very clear 
signature for fast orbital motion of the two sub-clusters. 
This motion results in a pronounced X-ray surface brightness gradient
in the southwestern part of the northern sub-cluster, and a highly elongated 
ICM surface brightness distribution of the southern sub-cluster. 
Although the merger has deformed the sub-clusters, their cores have not 
yet disintegrated. As discussed extensively in recent literature, a substantial 
magnetic field may contribute to the stability of 
cold and dense regions \cite{Vikh:01,Hein:03,Okab:03,Asai:04}.

Although the X-ray morphology suggests the existence of shocks towards the southwest
of A~115~N and a Mach cone towards the northeast, the
X-ray spectroscopy data are not consistent with the shock interpretation.
We conclude that the northern sub-cluster moves sub-sonically through the
ambient medium. The surface brightness jump in the southwest is caused by
a contact discontinuity of dense material moving through thinner material.
The bright X-ray emission in the northwest is caused by a fan of ICM left 
behind by the advancing sub-cluster core.
We find hot ICM in between the two sub-clusters, but no direct evidence for 
shock heating of the ICM. A possible interpretation is that the ICM is heated 
by viscous dissipation, rather than by shocks.

Estimates show that mergers with large impact parameters are usually 
associated with three or more mass concentrations \cite{Sara:02}.
Given that A~115~N is more massive and the presence of the large
distorted radio relic in the northeast of the sub-cluster, we speculate that it picked up
angular momentum during an earlier merger event.
It is instructive to compute the sub-cluster velocities, assuming that they 
are in a circular Keplerian orbit. We use the sub-cluster masses of the northern 
and southern sub-clusters of $m_1\,=$ 179$\times 10^{12}$ M$_\odot$ and 
$m_2\,=$ 63.9$\times 10^{12}$ M$_\odot$ from a surface brightness deprojection 
analysis \cite{Whit:97}. 
Assuming that the orbital plane is perpendicular to the line of sight, Kepler's third law
gives an orbital period

\begin{equation}
P\,=\,\left(\frac{4\, \pi^2\,a^3}{G\,(m_1+m_2)}\right)^{1/2}
\end{equation}

With a semi-major axis of $a\,=$ 484 kpc (2.5') and the gravitational constant
$G\,=$ 6.673$\times$ 10$^{-11}$ m$^3$ kg$^{-1}$ s$^{-2}$, we get 
$P\,=$ 2 billion years. The period corresponds to orbital velocities 
for A~115~N and A~115~S of 776 km s$^{-1}$ and 2147 km s$^{-1}$, 
respectively. While the velocity of A~115~N lies below the 
sound speed of the ambient medium $\sqrt{T/\rm 5\,keV}$ 1160 km s$^{-1}$, 
that of A~115~S exceeds the sound speed.
The velocities would be higher if the sub-clusters are in an elliptical or 
open orbit, and lower if the orbital plane is not seen edge-on.
If projection effects are negligible, the absence of a bow-shock surrounding 
the southern sub-cluster could be explained in two ways: (i) the ICM 
surrounding the sub-cluster moves with a similar velocity as the sub-cluster; 
(ii) the velocity of the sub-cluster is substantially below the one computed
for a Keplerian orbit as much of the kinetic energy has been dissipated into 
random energy of the ICM and the dark matter. 
\subsection{The core of the northern sub-cluster}
We found a compact cold core at the center of the northern sub-cluster with a projected 
width and height of 10'' (30 kpc) each.
A significant surface brightness discontinuity separates the core from the ambient ICM. 
The 1.4 GHz picture shows that the jets from the radio source 3C~28.0 
have drilled 2 channels through the X-ray bright core region and transition into radio lobes
upon leaving the core region. We find 3 viable interpretations:
\begin{enumerate}
\item The compact core might be the ISM of the central cD galaxy. 
See Kazuo et al.\ (2001) for a related discussion.
This interpretation might explain the surface brightness jump 
associated with this region by cold and dense ISM accumulating 
near the core of the cluster.
\item We may see the core of the cooling flow. The surface brightness
discontinuities might be produced by ``sloshing'' of the core, as recently reported for a 
large number of galaxy clusters observed with Chandra \cite{Mark:02}.
\item  Wilson et al.\ (2002) discovered filaments of cold material inside the
X-ray cocoon of the radio galaxy Cygnus-A. They interpreted this material as 
low entropy ICM sinking toward the core of the radio galaxy, and ultimately
fueling the activity of the AGN. Naively one would not expect
that this process produces ICM contact discontinuities, and thus drastic 
surface brightness jumps. However, the Chandra image of  Cygnus-A clearly shows 
strong surface brightness and temperature gradients throughout the central region 
of the cocoon-like structure associated with the radio source.
\end{enumerate}
The structure of the cores of galaxy clusters and the implications for fueling 
of the central AGN are clearly extremely important open questions.

%
%

As described in the introduction, heating of the ICM by the jets from the
AGN of the central cD galaxy is being discussed as a mechanism to explain
the absence of cold gas ($k_{\rm b}\,T$ $<$ 1.5 keV) in cooling flow
clusters. In the case of the radio galaxy Cygnus-A, Wilson et~al.~(2002)
found some evidence that the central radio galaxy heats the ICM, either
through a bow-shock or through p$\,\cdot\,$dV work done by the inflating radio
plasma on the ambient ICM.  The radio galaxy 3C 28.0 fits neatly into an
elliptical region of bright ICM, delimited to the southwest by the S1/S2
surface brightness jump. The cocoon-like morphology suggests that the
radio galaxy may drive a bow-shock into the ICM heating it.
However, we do not find spectroscopic evidence that bolsters this
hypothesis. In the case of Cygnus-A, the Chandra data shows that the
surface brightness and temperature varies on small spatial scales. The
poorer signal to noise ratio of the A~115 observations does not allow us
to search for similar small scale features.

\begin{figure}
\plotone{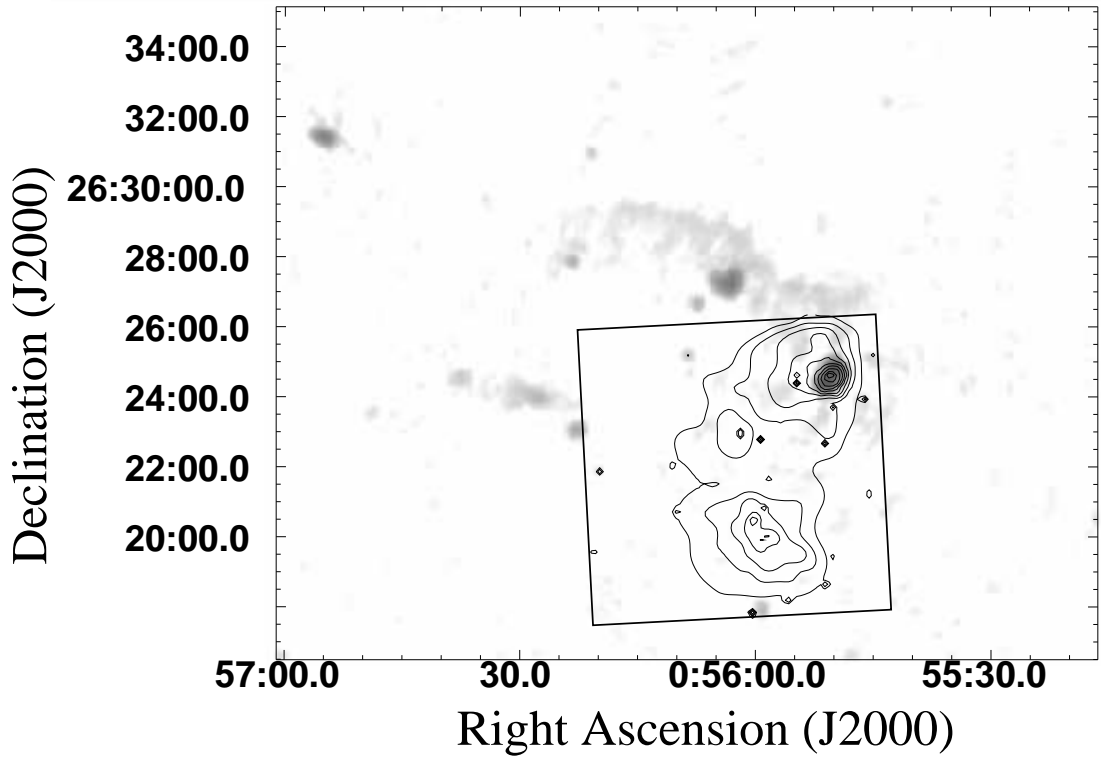}
\caption{Deep 1.4 GHz radio map of A~115 (grayscale) and Chandra X-ray surface brightness contours  to aid the eye.
The radio source 3C~28.0 can be recognized at the center of the northern sub-cluster peaking at 
0.96 Jy beam$^{-1}$\cite{Govo:01}.
An additional source peaking at 0.17 Jy beam$^{-1}$, and a faint arc of radio emission, radiating
at $\sim$2 mJy beam$^{-1}$ can be seen towards in northeast of the
cluster.  Beam size is $35'' \times 35''$ at FWHM.  This image spans 19.3' $\times$
24.3' (3.74 Mpc $\times$ 4.71 Mpc).}

\label{Gutierrez:RadioHalo}
\end{figure}

\acknowledgements
{\it Acknowledgments:} L. Govoni kindly sent us the deep 1.4 GHz radio data shown in Fig.\ 
\ref{Gutierrez:RadioHalo}. 
We obtained the radio map of 3C~28.0 from the 3C DRAGN atlas, 
updated and compiled by J.~ P.~ Leahy, A.~ H.~ Bridle, and R.~ G.~ Strom, {\it An Atlas of DRAGNs}.  
We thank M. Markevitch for prompt technical advice, whenever needed.
K.~G.~ and H.~K.~ acknowledge support by NASA/Smithsonian grant G02-3182X

\clearpage

\clearpage

\begin{table}
\begin{center}
\begin{scriptsize}
\begin{tabular} {|p{1.0cm}|c|c|c|c|c|}\hline
{\it Lable} & {\it Counterpart} & {\it Object type} & {\it RA/DEC} & {redshift} &{\it net counts}
\\ \hline \hline
{A} & {ZHG90/ACO 115 4} &  {cluster galaxy/radio source} & 00:56:00.3 / +26:17:50.3 & {--} & 107
\\ \hline
{B} & {BHG83(G 29)} & {cluster galaxy} &  00:55:00.2 / +26:20:32  & 0.192043 & 112
\\ \hline
{C} & {BHG83(G 28)} & {cluster galaxy} &  00:55:59.0 / +26:20:49.4 & 0.20282 & 68
\\ \hline
{D} & {BHG83(G 16)} & {cluster galaxy} & 00:56:19.7 / +26:21:52.1 & 0.190675 & 85
\\ \hline
{E} & {IRAS F00536+2611} & {Weak Infrared source} & 00:56:20.1 / +26:28:13.9 & {--} & 82
\\ \hline
{F} & {CXOU J005509.2+262714} & {No Named Counterpart} & 00:55:09.2 / +26:27:14 & {--} & 948
\\ \hline
\end{tabular}
\end{scriptsize}
\end{center}
\caption{ Localized X-ray sources discovered in the Chandra field of view.  
Sources A-D can be seen in Fig.~\ref{Gutierrez:region}. The Net Chandra counts are in the 0.5 keV to 6 keV energy band. }
\label{Gutierrez:t2}
\end{table}

\begin{table}
\begin{center}
\begin{tabular} {|p{3.9cm}|c|c|c|c|}\hline
{\it Region} & {\it Temperature (keV)} & {\it n$_{H}$} & {\it Metallicity} & {\it $\chi^{2}$(dof)} 
\\ \hline 
{N.~ Sub-Cluster} & $3.43^{+0.09}_{-0.09}$ &  $0.067^{+0.005}_{-0.005}$ & $0.37^{+0.03}_{-0.03}$ & $82.3(70)$
\\ \hline
{N.~ Sub-Cluster Core} & $2.19^{+0.08}_{-0.08}$ & $0.054^{+0.008}_{-0.010}$ & $0.44^{+0.03}_{-0.03}$ & $59.8(48)$
\\ \hline
{N.~ Sub-Cluster Fan} & $5.05^{+0.41}_{-0.39}$ & $0.054^{+0.010}_{-0.013}$ &  $0.19^{+0.07}_{-0.07}$ & $11.0(15)$
\\ \hline
{S.~ Sub-Cluster} & $5.35^{+0.47}_{-0.46}$ & $0.054^{+0.014}_{-0.015}$ & $0.19^{+0.09}_{-0.08}$ & $14.7(12)$
\\ \hline
{S.~ Cool Region} & $3.65^{+0.21}_{-0.16}$ & $0.076^{+0.012}_{-0.012}$ & $0.22^{+0.06}_{-0.06}$ & $14.4(17)$
\\ \hline
{Intermediate Region} & $8.70^{+1.50}_{-1.22}$ & $0.054^{+0.009}_{-0.022}$ & $0.013^{+0.111}_{-0.107}$ & $37.5(32)$
\\ \hline
{Connecting Region} & $7.19^{+0.53}_{-0.46}$ & $0.072^{+0.008}_{-0.08}$ & $0.22^{+0.06}_{-0.06}$ & $62.9(57)$
\\ \hline
\end{tabular}
\end{center}
\caption{Projected Plasma Parameters (Raymond Smith).}
\label{Gutierrez:t1}
\end{table}

\end{document}